\pdfoutput=1
\documentclass[12pt,letterpaper,floatfix]{article}
\usepackage{amssymb}
\usepackage{hyperref}
\usepackage{cite}
\usepackage{amsmath,amssymb,amsthm,bbm,color}
\usepackage{graphicx}
\usepackage{epsfig}
\usepackage{euscript}
\usepackage{pslatex}
\usepackage{fancybox}
\usepackage{enumerate}
\usepackage{widetext}
\usepackage{subfigure}
\usepackage{wasysym}
\usepackage{orcidlink}
\usepackage[compat=1.0.0]{tikz-feynman}

\definecolor{brightturquoise}{rgb}{0.03, 0.91, 0.87}
\definecolor{awesome}{rgb}{1.0, 0.13, 0.32}
\definecolor{armygreen}{rgb}{0.29, 0.33, 0.13}
\definecolor{aqua}{rgb}{0.0, 1.0,1.0}
\definecolor{maroon(html/css)}{rgb}{0.5, 0.0,0.0}
\definecolor{pinegreen}{rgb}{0.0, 0.47,0.44}
\definecolor{red-brown}{rgb}{0.65, 0.16,0.16}

\usepackage[figuresright]{rotating}

\begin{document}
\begin{titlepage}

\begin{center}
\begin{flushleft}
{\small \bf BU-HEPP-24-07, Nov 2024}
\end{flushleft}
\vspace{18mm}

{\bf Dirac-Schwinger Quantization for Emergent Magnetic Monopoles?}\\
\vspace{2mm}

{A. Farhan$^a$$^{\orcidlink{0000-0002-2384-2249}}$,~M. Saccone$^b$$^{\orcidlink{0000-0001-7400-9081}}$,~B.F.L. Ward$^a$$^{\orcidlink{0000-0003-0716-7850}}$}\\
{$^a$Department of Physics and Astronomy, Baylor University, Waco, TX, USA}\\
{$^b$Croputation, Santa Fe, NM, USA}\\
\end{center}
\centerline{\bf Abstract}
In Refs.~\cite{Dirac:1931,Dirac:1948,Schwinger:1968,Schwinger:1975} Dirac and Schwinger showed the existence of a magnetic monopole
required a charge quantization condition which we write following Dirac as $\frac{eg}{4\pi\hbar}=\frac{n}{2},\; n=0,\pm 1,\; \pm 2, \ldots$. Here, $g$ is the magnetic monopole charge and $e$ is the electric charge of the positron. Recently, in Refs. ~\cite{Farhan:2016,Farhan:2019}, it has been shown experimentally that frustrated spin-ice systems exhibit 'emergent' magnetic monopoles. We show that, within the experimental errors, the respective magnetic charges obey the Dirac-Schwinger quantization condition. Possible implications are discussed.
\end{titlepage}

Dirac~\cite{Dirac:1931,Dirac:1948} and Schwinger~\cite{Schwinger:1968,Schwinger:1975} have shown that quantum mechanics requires that any observable particle with a magnetic charge, $g\equiv g_J$, must obey the charge quantization condition 
\begin{equation}
\frac{eg}{4\pi\hbar} \equiv \frac{eg_J}{4\pi\hbar}=\frac{n}{2},\; n=0,\pm 1,\; \pm 2, \ldots,
\label{eqn-1}
\end{equation}
where e is the electric charge of the positron. Here, we follow Ref.~\cite{Jackson-3rdEd} in the definition of the monopole charge $g_J$. There have been many efforts to search for an isolated fundamental magnetic monopole with no success~\cite{mnple-srch-abbasi:2022}. Recently, in an unexpected development,
there has been a change in the experimental situation.\par
Specifically, in Refs.~\cite{Farhan:2016,Farhan:2019,Harris1997,Castelnovo2008}, the phenomenon of emergent magnetic monopoles has been observed in frustrated spin-ice systems, in naturally occurring pyrochlore spin ice~\cite{Harris1997,Castelnovo2008} or artificial spin ice systems, comprising Ising-type single-domain nanomagnets lithographically defined onto offset square lattices~\cite{Perrin2016,Farhan:2019}. In particular, the latter type of systems allowed for the direct real-space observation of emergent magnetic monopole dynamics as a function of time and temperature. We refer the reader to Ref.~\cite{Farhan:2019} for further details of such systems and the corresponding measurements thereon. Here, we want to investigate the applicability of the Dirac-Schwinger (DS) quantization condition to the observed emergent magnetic monopoles.\par 
In Refs.~\cite{Farhan:2016,Farhan:2019}, emergent magnetic monopoles whose magnetic charges
can be seen to be consistent with the DS quantization condition were observed within the errors as follows.
Employing the so-called dumbbell model~\cite{Castelnovo2008,Moller2009}, magnetic moments of the nanomagnets are replaced by pairs of magnetic charges $\pm g_{CMS}$ residing at the tips of the nanomagnets. The magnitude of a charge $g_{CMS}\equiv g_J/\mu$ is given by the magnetic moment divided by the length of the nanomagnet. Magnetic configurations that constitute defects in the form of ice-rule violations then act as defects that carry magnetic charges of $2g_{CMS}$. The temperature-dependent motion of these defects is then observed via real-space magnetic imaging and directly compared to a model describing weakly-interacting plasmas~\cite{Levin2002} of magnetic charges emerging through Bjerrum ion pairing. An agreement between the model and experiments was achieved for a monopole charge of $Q = 2g_{CMS} = 9.765\times10^{-12} \;Am$.  
Translating this value for the magnetic charge in the dumbbell model of Ref.~\cite{Castelnovo2008} to the definition used in Eq.(\ref{eqn-1}), we instate a factor of the permeability of the permalloy $\mu\equiv \mu_r\mu_0$, with $\mu_r \cong (9.0\pm 0.2)\times 10^4$ following Ref.~\cite{Sun:2023}.
Employing this monopole charge into the DS quantization condition mentioned above yields the results
\begin{equation}
\begin{split}
g_J&= \mu_r\mu_0\times 9.765\times10^{-12}Am = 1.1044\times 10^{-12} J/A, \\
g_Je/(4\pi \hbar) &= 1.1044\times 10^{-12} (J/A)\times 1.602177\times 10^{-19}C/(4\pi \cdot 1.054572\times 10^{-34} Js) \\
&= 133.5 \pm 13.7.
\end{split}
\label{eqn-2}
\end{equation}
In the last equation, the error has been estimated from the errors $\Delta Q/Q \cong \pm 0.1$ and $\Delta \mu_r/\mu_r \cong \pm 0.022$ ~\cite{Sun:2023}.
These results support that, within errors, quantum mechanics is respected by the corresponding emergent phenomena. Indeed, we note that our results eliminate the possibility that the LHS of the second equation in eqs. (2) is too small to be an integer in magnitude at the level of $9.67\;\sigma$ significance. This is a new result as is the measurement of the respective LHS value to 10\% accuracy. \par
From a purely theoretical perspective, one expects these emergent monopoles to respect the DS quantization condition because the arguments underlying the condition involve (Dirac formulation) closed-line integrals for the respective vector potentials and are
only dependent on the changes in the solid angle of the gauge difference surface at the charge as it crosses that surface or involve (Schwinger formulation) gauge difference four-volume integrals and are only dependent on changes in the action contribution as the charge enters the attendant action gauge difference four-volume -- the arguments do not depend on the details of the magnetic charges which engender the respective gauge transformations. This has suggestions that, in the theory of elementary particles, a frustrated spin-ice type scenario, at a scale well below the Fermi scale, could also realize emergent magnetic monopoles? Thus, we encourage the corresponding continued search for them.\par
We note that, in Ref.~\cite{Castelnovo2008}, the possibility of applying the DS condition in the pure spin-ice system not organized in nanomagnets as in Ref.~\cite{Farhan:2019} was discussed and dismissed. The assertion in Ref.~\cite{Castelnovo2008} that there are no Dirac strings that lead to the DS quantization condition is incorrect. The Dirac string formulation of the monopole fields is always alive, even in the case that a series of dipoles connects the two monopoles of opposite charge. As Dirac~\cite{Dirac:1948} has emphasized, his strings are unobservable and admit changes in them via gauge transformations,
which always lead to the DS quantization condition. Similarly, the Schwinger~\cite{Schwinger:1975} formulation also still holds, as it only depends on the local entry of the charge into the respective gauge difference four-volume.\par

\vskip 2 mm
\vskip 2 mm


\begin{thebibliography}{99}
\bibitem{Dirac:1931} P. A. M. Dirac, Proc. R. Soc. London Ser. A {\bf 133} (1931) 60.
\bibitem{Dirac:1948} P. A. M. Dirac, Phys. Rev. {\bf 74} (1948) 817.
\bibitem{Schwinger:1968} J. Schwinger, Phys. Rev. {\bf 173} (1968) 1536.
\bibitem{Schwinger:1975} J. Schwinger, Phys. Rev. D {\bf 12} (1975) 3105.
\bibitem{Farhan:2016} A. Farhan {\it et al.}, Nature Communications {\bf 7} (2016) 12635.
\bibitem{Farhan:2019} A. Farhan {\it et al.}, Science Advances {\bf 5} (2019) eaav6380.
\bibitem{Jackson-3rdEd} J.D. Jackson, {\it Classical Electrodynamics, 3rd Ed.} (John Wiley \& Sons, Inc., Hoboken, NJ, USA, 1999).
\bibitem{mnple-srch-abbasi:2022} R. Abbasi et al., Phys. Rev. Lett. 128 (2022) 051101, and references therein.
\bibitem{Harris1997} M. J. Harris, S. T. Bramwell, D. F. McMorrow, T. Zeiske, and K. W. Godfrey, Phys. Rev. Lett. {\bf 79} (1997) 2554.
\bibitem{Castelnovo2008} C. Castelnovo, R. Moessner, and S. L. Sondhi, Nature {\bf 451} (2008) 42 – 45.
\bibitem{Perrin2016} Y. Perrin, D. Canals, and N. Rougemaille, Nature {\bf 540} (2016) 410 -- 415.
\bibitem{Moller2009} G. Möller and R. Moessner, Physical Review B {\bf 80} (2009) 140409(R).
\bibitem{Levin2002} Y. Levin, Rep. Prog. Phys. {\bf 65} (2002) 1577.
\bibitem{Sun:2023} J. Sun {\it et al.}, Materials (Basel) {\bf 16} (2023) 3253.
\end{thebibliography}

\end{document}